\documentclass[conference]{IEEEtran}
\IEEEoverridecommandlockouts
\usepackage{cite}
\usepackage{amsmath,amssymb,amsfonts}
\usepackage{graphicx}
\usepackage{textcomp}
\usepackage{xcolor}
\usepackage{booktabs}
\usepackage{multirow}
\usepackage{amsthm}
\usepackage{tcolorbox}
\usepackage{adjustbox}
\usepackage{makecell}
\usepackage{enumitem}
\usepackage{svg}
\usepackage{tikz}
\usepackage{algorithm} 
\usepackage{algorithmic}
\usepackage[utf8]{inputenc}
\usepackage{pifont} 
\usepackage{array}
\usepackage{ragged2e}
\usepackage{xurl}
\usepackage{balance}
\usepackage{hyperref}

\DeclareUnicodeCharacter{2713}{\ding{51}} 
\DeclareUnicodeCharacter{2717}{\ding{55}} 

\definecolor{cosmeticGreen}{RGB}{204, 219, 152}
\definecolor{structureYellow}{RGB}{253, 231, 153}
\definecolor{semanticPink}{RGB}{245, 220, 225}

\DeclareRobustCommand{\baseBlock}[4]{%
    \ifmmode\text{\tikz[baseline=(X.base), inner sep=0pt, outer sep=0pt] \node[...];}\else
    \tikz[baseline=(X.base), inner sep=0pt, outer sep=0pt] \node[%
        fill=#1,                
        draw=#2,                
        #3, 
        line width=0.6pt,       
        rounded corners=2pt,  
        inner sep=1.2pt,       
        anchor=base,         
        font=\small
    ] (X) {#4};%
    \fi
}

\newcommand{\scopeCosmetic}[1]{\baseBlock{cosmeticGreen}{none}{solid}{#1}}
\newcommand{\scopeStructural}[1]{\baseBlock{structureYellow}{none}{solid}{#1}}
\newcommand{\scopeSemantic}[1]{\baseBlock{semanticPink}{none}{solid}{#1}}

\newcommand{\changeAdd}[1]{\baseBlock{none}{black}{solid}{\textbf{#1}}}
\newcommand{\changeRemove}[1]{\baseBlock{none}{black}{dashed}{\textbf{#1}}}
\newcommand{\changeModify}[1]{\baseBlock{none}{black}{dash dot}{\textbf{#1}}}

\newcommand{\CosmeticRemove}[1]{\baseBlock{cosmeticGreen}{cosmeticGreen!85!black}{dashed}{#1}}

\theoremstyle{definition}
\newtheorem{definition}{Definition}[section]

\newcommand{\proc}[1]{\textsc{#1}} 
\newcommand{\kw}[1]{\textbf{#1}}
\newcommand{\mycomment}[1]{\hfill \textcolor{darkgray}{\textit{\footnotesize \# #1}}}

\def\BibTeX{{\rm B\kern-.05em{\sc i\kern-.025em b}\kern-.08em
    T\kern-.1667em\lower.7ex\hbox{E}\kern-.125emX}}

\begin{document}
\title{CodeChat-Eval: Evaluating Large Language Models in Multi-Turn Code Refinement Dialogues}


\author{
    \IEEEauthorblockN{
        Guoxiang (Aaron) Guo\IEEEauthorrefmark{1},
        Kla Tantithamthavorn\IEEEauthorrefmark{1},
        Neelofar Neelofar\IEEEauthorrefmark{2}, \\
        Yuanyuan Qi\IEEEauthorrefmark{1}, \textit{and}
        Aldeida Aleti\IEEEauthorrefmark{1}
    }
    \vspace{0.3cm}
    \IEEEauthorblockA{\IEEEauthorrefmark{1}\textit{Faculty of Information Technology}, \textit{Monash University}, Clayton, VIC, Australia \\
    Email: \{Guoxiang.Guo, Chakkrit, Yuanyuan.Qi, Aldeida.Aleti\}@monash.edu}
    \vspace{0.1cm}
    \IEEEauthorblockA{\IEEEauthorrefmark{2}\textit{School of Computing Technologies}, \textit{RMIT University}, Melbourne, VIC, Australia \\
    Email: neelofar.neelofar@rmit.edu.au}
}

\maketitle
\begin{abstract}
Large Language Models (LLMs) are increasingly used in software engineering to generate and refine code. In practice, developers often continue from an initial code generation request with follow-up refinement instructions, such as requests to improve style, restructure implementation, or change the execution strategy while preserving the intended behaviour. However, existing benchmarks generally omit this multi-turn code refinement dialogue setting and therefore cannot evaluate whether LLMs maintain functional correctness, i.e., whether the refined code still passes the test suite for the original task. To address this limitation, we introduce CodeChat-Eval, an evaluation framework that constructs evaluation sessions from multi-turn code refinement dialogues using a dynamic instruction selection algorithm. Our empirical study on open-weight and proprietary LLMs observes a statistically significant decrease ranging from 19.2\% (GPT-5 Nano) to 69.2\% (Llama 3.1 8B) in functional correctness over multi-turn refinement. The largest correctness drops are associated with logic-level refinements and additive change requests. These findings indicate that LLMs struggle to maintain functional correctness during multi-turn code refinement dialogues, and highlight the need for benchmarks that evaluate functionality-preserving refinement beyond single-turn generation.
\end{abstract}

\section{Introduction}

Large Language Models (LLMs) are increasingly integrated into real-world software development workflows, supporting both code generation and refinement~\cite{danyaro2025llm}. These interactions often unfold over multiple turns, as developers iteratively refine generated code through follow-up instructions~\cite{zhong2025developerLLMconv}. A typical dialogue begins with an initial code-generation request (e.g., \emph{``write a Python function to remove odd numbers from a given list''}), followed by refinement instructions that adapt the code to individual preferences or team coding standards (e.g., \emph{``remove all comments''}).

\begin{figure}[!t]
    \centering
    \includegraphics[width=\linewidth]{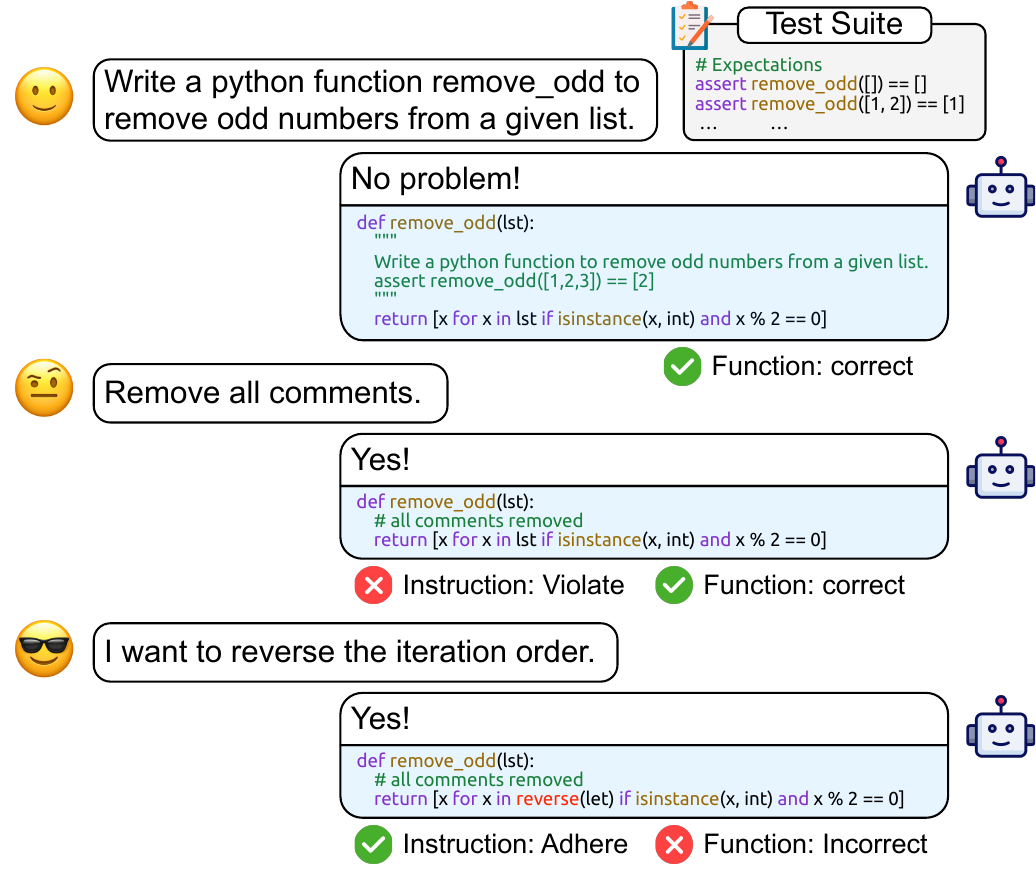}
    \caption{A motivating example of multi-turn code refinement that demonstrates the scenarios when LLMs fail to adhere to a developer's instruction or fail to maintain the functional correctness (i.e., do not pass the test suite).}
    \label{fig:2_3_exp}
\end{figure}

Ideally, such LLMs must revise the code as per the developer-preferred instructions (defined as \emph{instruction adherence}), while maintaining the \emph{functional correctness} of the code (i.e., the generated code fully passes the given test suite)~\cite{liu2024empiricalllmrefactor, zhong2025developerLLMconv, xiao2024devgpt, mohamed2025impactllmcode}.
However, in follow-up turns, there are many cases that LLMs may fail to adhere to the developer's instruction (i.e., a given instruction is not applied in the follow-up turn) or fail to maintain the functional correctness.
As illustrated in our motivating example (see Fig.~\ref{fig:2_3_exp}), the LLM already fails to adhere to the developer's instruction in the second turn (\emph{``remove all comments"} - the comment is not fully removed), and fails to maintain the functional correctness (``\emph{reverse the iteration order"} - the iteration order is reversed, but the LLM-generated code does not fully pass the test suite).

Recently, researchers introduced various software engineering (SWE) benchmark datasets to evaluate the coding capabilities of LLMs.
In particular, such datasets are designed to measure the functional correctness of the LLM-generated code in a single-turn scenario, such as MBPP\cite{MBPP}, HumanEval\cite{humaneval}, SWE-Bench\cite{jimenez2023swebench}, and BigCodeBench\cite{bigcodebench}, without considering whether LLM-generated code remains functionally correct across multi-turn code refinement.
Thus, little is known about how well the LLMs can maintain the functional correctness and adhere to the developers' instructions in multi-turn code refinement, and what types of developers' instructions have the largest impact on the functional correctness of the LLM-generated code.
Knowing how well LLMs maintain correctness and adhere to different types of developer instructions helps developers understand the strengths and weaknesses of LLMs for multi-turn code refinement capabilities beyond the single-turn code generation and decide which LLMs should be adopted in the practical software engineering workflows.

In this paper, we introduce CodeChat-Eval, a dynamic evaluation framework for multi-turn code refinement.
To set up an initial code generation turn, we used 164 SWE tasks from HumanEval~\cite{humaneval} and 378 SWE tasks from MBPP~\cite{MBPP}.
Each task consists of a pair of $<$instruction, test suite$>$, where the instruction is used for the initial code generation, and the test suite is used as a proxy to determine the functional correctness of the LLM-generated code.
To generate follow-up code refinement turns, our framework consists of three steps: instruction curation; evaluation agenda; and dynamic instruction selection.
First, we develop a taxonomy of code refinement instructions.
Our taxonomy characterises each refinement instruction by its \emph{scope}, i.e., the depth of intervention required by the instruction (\scopeCosmetic{cosmetic}, \scopeStructural{structural}, or \scopeSemantic{semantic}), and its \emph{change operation}, i.e., whether the instruction asks the model to \changeAdd{add}, \changeRemove{remove}, or \changeModify{modify} code content or patterns. These follow-up instructions are curated from the CodeAlignBench dataset~\cite{mehralian2025codealignbench}.
Then, our framework creates a pre-defined evaluation agenda (i.e., the sequence of the code refinement types for the $n$-turn evaluation session).
Then, we introduce an agenda-guided dynamic instruction selection algorithm, where this algorithm will find a code refinement instruction that is applicable and realistic to a prior version of the code (e.g., \CosmeticRemove{remove comment} should not be applied if the prior version that does not have comments). 

Finally, for each LLM-generated code snippet at each turn, we measure functional correctness (i.e., whether the code passes the given test suite) and instruction adherence (i.e., whether the requested refinement is actually applied, judged using an LLM-as-a-Judge).
Through an empirical evaluation of 542 SWE tasks over 10-turn evaluation sessions on eight open-weight and proprietary LLMs from four families (i.e., Llama, Qwen, DeepSeek, and GPT), we answer the following two research questions.

\begin{enumerate}[label=\textbf{RQ\arabic*})]

\item \textbf{How well do LLMs maintain functional correctness when applying the code refinement instructions?}\\
We find that the functional correctness of the LLM-generated code decreases by 19.2\% (GPT-5 Nano) to 69.2\% (Llama 3.1 8B) after applying follow-up refinement instructions. This decline is statistically significant across all evaluated LLMs, indicating that they struggle to maintain functional correctness in multi-turn code refinement.

\item \textbf{Which types of code refinement instructions have the largest impact on the functional correctness of the LLM-generated code?} \\
We find that \scopeSemantic{Semantic} (decreased the functional correctness by 21\%) instructions and \changeAdd{Add} (decreased the functional correctness by 17\%) instructions have the largest negative impact on the functional correctness of the LLM-generated code.

\end{enumerate}

\section{Related Work \& Research Questions}
In this section, we discuss the related work in order to formulate the research questions.

Large Language Model (LLM)-powered coding assistants, such as Claude Code\cite{anthropic_claude_code_2025_misc} and GitHub Copilot\cite{GitHubCopilot}, are increasingly used in software development workflows where developers iteratively request code generation and refinement. 
This interaction pattern makes code generation an incremental maintenance activity: each follow-up instruction may improve style, restructure implementation, or alter the execution strategy while the original function interface and behaviour should remain intact. 
Therefore, evaluating LLM-based coding assistants requires benchmarks that measure whether code remains correct after developer-preferred refinement instructions, not only whether the first generated solution passes a test suite.

\subsection{Functional Correctness in Code Generation Benchmarks}
Task Success Rate is an evaluation objective used to evaluate how well an LLM successfully solves a SWE task, which is widely used in existing SWE benchmark datasets (e.g., HumanEval\cite{humaneval}, MBPP\cite{MBPP}, SWE-Bench\cite{jimenez2023swebench}, and BigCodeBench\cite{bigcodebench}).
To operationalize task success, prior studies commonly use Pass@$k$ as a proxy to measure the percentage of SWE tasks that fully pass the test suite within $k$ attempts.
However, these benchmarks predominantly focus on single-turn code generation or issue resolution.
Even when an evaluation involves multiple internal steps, such as the agent workflow in SWE-Bench, the user-facing task remains a single fixed request rather than an externally driven sequence of refinement instructions.
As a result, these benchmarks do not evaluate whether LLMs can preserve functional correctness when developer intent is expressed through follow-up code refinement instructions.
Therefore, we formulate the following research question:

\begin{tcolorbox}[colback=gray!0, colframe=black, boxrule=0.5pt, top=1mm, bottom=1mm, left=1.5mm, right=1.5mm, boxsep=0pt, before skip=1.5mm, after skip=1.5mm]
\textbf{RQ1: How well do LLMs maintain functional correctness when applying the code refinement instructions?}
\end{tcolorbox}

\subsection{Multi-Turn Code Refinement and Instruction Adherence}
Recent benchmarks have started to examine LLM behaviour beyond single-turn code generation.
MINT evaluates broad tool augmented and feedback-driven exchanges\cite{wang2024mint}, which are adjacent to but broader than software code refinement.
Code-oriented benchmarks such as CodeAlignBench\cite{mehralian2025codealignbench}, CodeIF\cite{yan2025codeif}, CodeIF-Bench\cite{wang2025codeifmt}, MultiCodeIF\cite{duan2025hierarchical}, and ConvCodeWorld\cite{han2025convcodeworld} evaluate developer-preferred code adjustments, instruction adherence, or conversational code generation.
However, existing work does not isolate functionality preserving follow-up code refinement and measure whether code that was previously correct becomes incorrect after refinement.
CodeChat-Eval therefore focuses on multi-turn code refinement: each follow-up turn asks the LLM to revise existing code while preserving the original function interface and behaviour.
Since different instruction scopes (such as \scopeCosmetic{cosmetic}, \scopeStructural{structural}, or \scopeSemantic{semantic} changes) and change operations (such as \changeAdd{add}, \changeRemove{remove}, or \changeModify{modify}) can affect functional correctness in different ways, it is important to identify which types of refinement instructions are most likely to trigger Pass-to-Fail regressions.
Therefore, we formulate the following research question:

\begin{tcolorbox}[colback=gray!0, colframe=black, boxrule=0.5pt, top=1mm, bottom=1mm, left=1.5mm, right=1.5mm, boxsep=0pt, before skip=1.5mm, after skip=1.5mm]
\textbf{RQ2: Which types of code refinement instructions have the largest impact on the functional correctness of the LLM-generated code?}
\end{tcolorbox}

Table~\ref{tab:related_work_comparison} contrasts these benchmarks by task, follow-up setting, and evaluation target.
CodeChat-Eval is the only benchmark that explicitly tracks whether functionally correct code regresses after functionality-preserving refinement.
\begin{table}[!t]
    \centering
    \caption{Comparison of representative benchmarks. IA denotes instruction adherence, FC denotes test-suite-based functional correctness, and FR denotes functional regression after multi-turn refinements.}
    \label{tab:related_work_comparison}
    \begin{adjustbox}{max width=\linewidth}
    \begin{tabular}{l|l|c|cc|c}
        \toprule
        \textbf{Benchmark} & \textbf{Primary Task} & \textbf{Follow-up} & \textbf{IA} & \textbf{FC} & \textbf{FR} \\
        \midrule
        HumanEval/MBPP \cite{humaneval,MBPP} & Code generation & -- & ✗ & ✓ & ✗ \\
        SWE-Bench \cite{jimenez2023swebench} & Issue resolution & -- & ✗ & ✓ & ✗ \\
        IFEval \cite{zhou2023ifeval} & General text & -- & ✓ & ✗ & ✗ \\
        CodeIF \cite{yan2025codeif} & Code generation & -- & ✓ & ✗ & ✗ \\
        MINT \cite{wang2024mint} & Tool-based tasks & Feedback & ✗ & -- & ✗ \\
        CodeIF-Bench \cite{wang2025codeifmt} & Code generation & Constraints & ✓ & ✗ & ✗ \\
        MultiCodeIF \cite{duan2025hierarchical} & Code generation & Correction & ✓ & ✗ & ✗ \\
        ConvCodeWorld \cite{han2025convcodeworld} & Code generation & Feedback & ✗ & ✓ & ✗ \\
        CodeAlignBench \cite{mehralian2025codealignbench} & Code adjustment & Refinement & ✓ & ✗ & ✗ \\
        \midrule
        \textbf{CodeChat-Eval} & \textbf{Code refinement} & \textbf{Refinement} & \textbf{✓} & \textbf{✓} & \textbf{✓} \\
        \bottomrule
    \end{tabular}
    \end{adjustbox}
\end{table}

\section{Methodology}
\label{sec:Meth}

\subsection{Overview}
\label{sec:3.0}
\begin{figure*}[!t]
    \centering
    \includegraphics[width=1\linewidth]{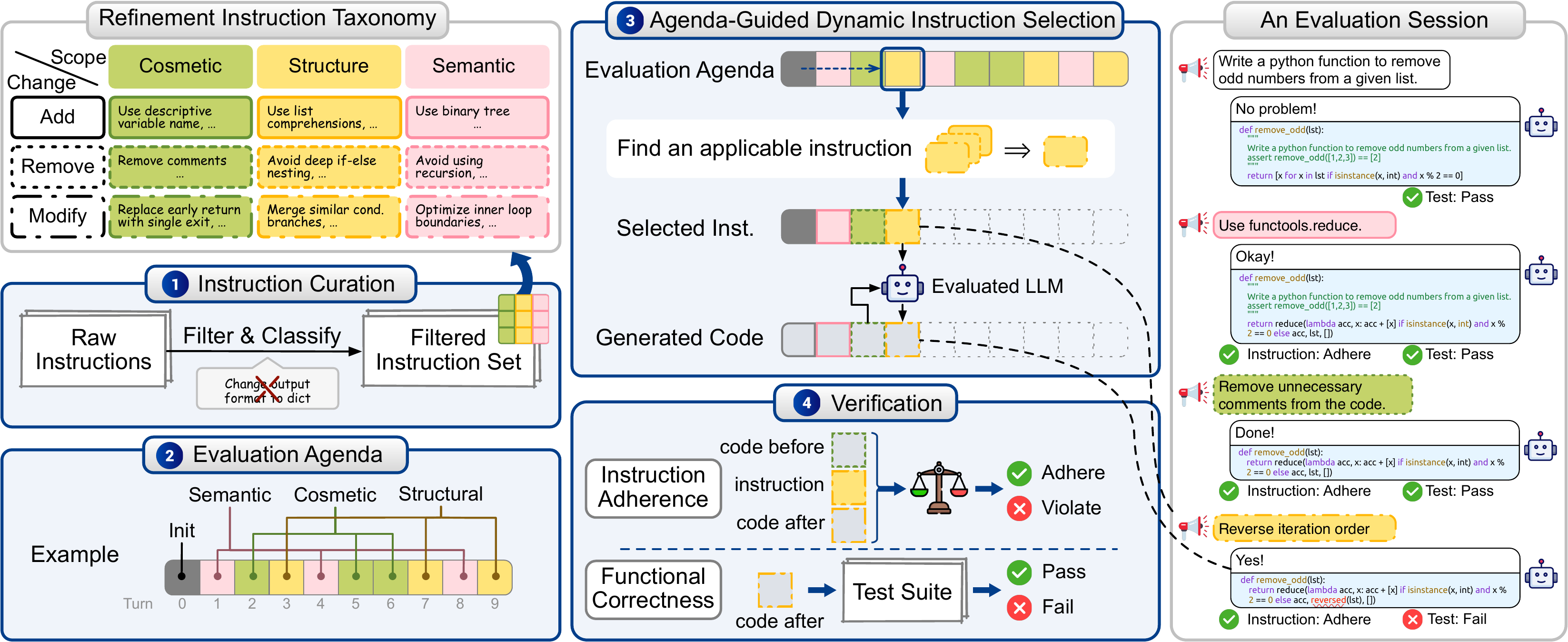}
    \caption{CodeChat-Eval overview.}
    \label{fig:3_1_overview}
\end{figure*}
CodeChat-Eval is an evaluation framework designed to assess LLMs in multi-turn code refinement dialogues. As shown in Fig.~\ref{fig:3_1_overview}, first, the raw instructions are cleaned and categorised into different types to form an instruction pool. The evaluation agenda is introduced to guide the generation of each evaluation session and achieves a balanced distribution of different types of instructions. Using dynamic instruction selection algorithm, CodeChat-Eval then adaptively selects the applicable refinement instruction as the input of the follow-up turn. Finally, the verification step evaluates functional correctness and records instruction adherence as a supporting measure, forming the whole evaluation session.

\subsection{Code Refinement Instructions}
\label{sec:3.1}

\subsubsection{The Instruction Taxonomy}
To understand the impact of different types of instructions, we propose a two-dimensional taxonomy shown in Fig.~\ref{fig:3_1_overview} to categorise code refinement instructions. The scope is defined as the depth of intervention required to fulfil the code refinement described in the instruction. Based on the the classification in CodeAlignBench\cite{mehralian2025codealignbench}, the different scopes are defined as follows:
\begin{itemize}
    \item \scopeCosmetic{Cosmetic}: modifications affecting readability, style, or presentation without altering the underlying logic. For example: add/remove comment, descriptive variable name.
    \item \scopeStructural{Structural}: modifications that restructure the code by applying a series of refactoring without changing its observable behaviour. For example: use list comprehensions, avoid deep nesting of if-else statements.
    \item \scopeSemantic{Semantic}: modifications that alter the underlying logic or execution profile of code while maintaining the same functionality and interface. For example: avoid using recursion, use binary tree.
\end{itemize}

In addition to the scopes, we further classify the instructions with different changes. The definitions of different changes are as follows:
\begin{itemize}
    \item \changeAdd{Add}: adding certain content, feature or pattern that is missing in the code to be revised. For example: add more comments, use list comprehension.
    \item \changeRemove{Remove}: removing certain content or features that exist in the code to be revised. For example: remove comments, avoid using recursion.
    \item \changeModify{Modify}: modifying certain content, feature, or settings existing in the code to be revised, with explicitly or implicitly mentioned alternative. For example: replace early return with single exit, merge similar conditional branches.
\end{itemize}

CodeChat-Eval uses this taxonomy to classify code refinement instructions. This classification enables CodeChat-Eval to evaluate the impact of different instruction scopes changes in multi-turn code refinement.

\subsubsection{Instruction Set Curation}
We adopt instructions collected from real-world developer-LLM dialogues\cite{mehralian2025codealignbench} as the raw instruction set. These instructions are verified to represent developer-aligned adjustments in readability, structure, or other functionality-preserving quality aspects. In CodeChat-Eval, the following steps are conducted to curate the instruction set:

\textbf{Instruction Classification:} we manually inspect and label each instruction based on their scope and change. The scope tags are derived from the original dataset\cite{mehralian2025codealignbench}. Any instruction that is ambiguous, contradictory, or has incomparable effects is excluded in CodeChat-Eval to ensure accurate judgement. The change tags are derived through a systematic lexical analysis. Since code refinement instructions use imperative verbs (e.g., ``\textit{Remove} comments''), CodeChat-Eval uses a keyword matching algorithm to map verbs to changes. As shown in Table~\ref{tab:3_2_keyword_mapping}, CodeChat-Eval maps 59 verbs to \changeAdd{Add}, \changeRemove{Remove}, and \changeModify{Modify} changes based on their operational semantics in software maintenance. The empirical results confirm that this rule-based approach achieved 100\% coverage across our curated instruction set. 

\begin{table}[h!] 
    \centering 
    \caption{Mapping from Keywords to Change Operations} 
    \label{tab:3_2_keyword_mapping} 
    \adjustbox{max width=\linewidth}{
    \begin{tabular}{l|l} 
        \toprule 
        \textbf{Change} & \textbf{Keywords} \\ 
        \midrule 
        \changeAdd{Add} & \textit{add, use, include, implement, ensure, define, annotate, ...} \\ 
        \changeRemove{Remove} & \textit{avoid, remove, eliminate, prohibit, prevent, minimise, ...} \\ 
        \changeModify{Modify} & \textit{replace, optimise, refactor, merge, split, convert, ...} \\ 
        \bottomrule 
    \end{tabular}} 
\end{table}

\textbf{Instruction Filtering:} During the inspection of instructions, we observe that certain code refinement instructions conflict with functional correctness evaluation harness. Modern functional correctness benchmarks, e.g., EvalPlus\cite{liu2023evalplus} and BigCodeBench\cite{bigcodebench}, rely on a rigid test harness that invokes code via fixed function signatures. Consequently, instructions that explicitly request modifying the function interface (e.g., ``Use a more descriptive function name.'') disrupt test runner, resulting in false negatives during execution (e.g., NameError). We elaborate on this phenomenon in Section~\ref{sec:3.4}. To address this issue, we manually filter out instructions that violate the interface consistency required by the evaluation framework.

In CodeChat-Eval, the number of available raw instructions is 169. Two authors independently label whether each raw instruction is functionality altering or conflicts with test harness. Later, the two authors resolved diverging labels through further discussion and exploratory experiments. In this process, 11 raw instructions (6.51\%) are filtered out. In the filtered instruction set, the most common scope is \scopeCosmetic{Cosmetic} (40\%), and the most common change is \changeAdd{Add} (50\%). CodeChat-Eval uses this filtered instruction set to evaluate LLMs in multi-turn code refinement.

\subsection{Evaluation Agenda}
\label{sec:3.2}
Instead of using purely random selection or fixed instruction sequence, the scopes of refinement instructions are predefined in a sequence in CodeChat-Eval. While feasible, purely random selection suffers from three key limitations: 1) different instruction types may not be evenly distributed in evaluation, 2) it introduces enlarged variance in difficulty\cite{laban2025llmsgetlost}, undermining reproducibility and fairness, and 3) random selection creates evaluation bias by over-representing specific scopes, making cross-model comparisons unreliable. To address these limitations, we propose the Evaluation Agenda.

\begin{definition}[Evaluation Agenda]
An evaluation agenda is a high-level arrangement that dictates the instruction type of each turn without fixing the specific instruction.
\end{definition}

Formally, let $\mathcal{T}$ denote the space of instruction types defined in our taxonomy. We define an agenda $\mathcal{A}$ as a parameterised sequence $\mathcal{A}(N, \mathcal{D}) = \langle \tau_0, \dots, \tau_{N-1} \rangle$, where $N$ is the number of total interaction turns and $\mathcal{D}$ represents the target distribution of instruction types over $\mathcal{T}$. For each turn $i \in [0, N-1]$, the instruction type $\tau_i$ is sampled from $\mathcal{D}$ (i.e., $\tau_i \sim \mathcal{D}$).

In CodeChat-Eval, we set the evaluation session length to 10 turns, covering 95\% of real-world scenarios \cite{zhong2025developerLLMconv}. Given that there are three different scopes in taxonomy ($\mathcal{T}=$\{\scopeCosmetic{Cosmetic}, \scopeStructural{Structural}, \scopeSemantic{Semantic}\}), we adopt a scope-balanced evaluation agenda comprising an initial turn and 9 follow-up turns (i.e., $\mathcal{D}$ assigns a uniform probability of 1/3 to each scope in $\mathcal{T}$). Therefore, the follow-up turns will consist of 3 \scopeCosmetic{Cosmetic}, 3 \scopeStructural{Structural}, and 3 \scopeSemantic{Semantic} refinement turns, and they are randomly ordered to increase diversity of evaluation sessions. Fig. \ref{fig:3_1_overview} illustrates an example agenda. The initial turn (Turn 0) requests a function implementation. In subsequent turns, the agenda specifies the scopes to guide the instruction selection.

Evaluation Agenda provides a feasible approach to guide evaluation session generation. Given the fact that the distribution of instructions used by real-world developers is inaccessible, the balanced Evaluation Agenda in CodeChat-Eval represents a best-effort approach to ensure fair evaluation.

\subsection{Evaluation Session Generation}
\label{sec:3.3}
CodeChat-Eval generates each evaluation session by integrating the evaluation agenda with in-turn instruction applicability checks. As illustrated in Fig.~\ref{fig:3_1_overview}, this pipeline takes the code generated in the previous turn and a pool of instructions as input. It then employs the \textbf{Agenda-Guided Dynamic Instruction Selection (AGDIS)} algorithm to select the code refinement instruction for the current turn. The AGDIS algorithm bridges the high-level strategy defined in the agenda and the state of the current code.

\subsubsection{Instruction Applicability Check}
\label{sec:3.3.1}

While the Evaluation Agenda defines the target scope (e.g., \scopeStructural{Structural}) for each turn, the specific instruction must be compatible with the current code. Inspired by the applicability check in CodeAlignBench\cite{mehralian2025codealignbench}, we implement a prompt-based \texttt{Is\_Applicable} function, which takes the code snippet and a piece of instruction as input, and outputs whether the instruction is applicable to the code snippet (\texttt{True}) or not (\texttt{False}). In the instruction applicability check (the function \texttt{Is\_Applicable}), Chain-of-Thought reasoning\cite{wei2022cot} is applied to improve the accuracy of output. The prompt template is provided in supplementary material. According to the prompt, the underlying reasoning LLM is required to think step-by-step and verify the conditions before producing the final verdict. Such process is fully automated. We further verify the accuracy of applicability check in Section~\ref{sec:4.5}.



\subsubsection{Session Construction}
An evaluation session records the interaction history between the simulated user and the evaluated LLM. Before the session starts, an agenda is generated to guide the evaluation session, and the entire filtered instruction set forms the instruction pool. As formalised in Algorithm~\ref{alg:agdis}, the AGDIS algorithm generates the session in three main phases:

\begin{enumerate}
    \item \textbf{Initialisation (Turn 0):} The session begins by prompting the LLM to solve the initial problem description, generating the initial code ($C_0$).
    \item \textbf{Refinement Loop (Turns 1-9):} For each follow-up turn $t$, the algorithm retrieves the target scope $\tau_t$ from the agenda $\mathcal{A}$. It then iterates through the instruction set corresponding to $\tau_t$, using the \texttt{Is\_Applicable} function to find a valid instruction.
    \item \textbf{Dynamic Selection \& Fallback:} Once an applicable instruction is identified, it is removed from the available pool to prevent repetition and sent to the LLM to generate the refined code ($C_t$). In the rare event that no applicable instruction is found, the algorithm records a \textit{SKIP} flag, maintaining the previous code state to continue.
\end{enumerate}

This process repeats until the evaluation session reaches the maximum number of turns ($N$), and generates the complete interaction history of the evaluation session.

\begin{algorithm}[t]
\caption{Agenda-Guided Dynamic Instruction Selection (AGDIS)}
\label{alg:agdis}
\begin{algorithmic}[1]
\small 
\REQUIRE Problem Statement $P$, Instruction Pool $\mathcal{I}$, Model $\mathcal{M}$, Agenda $\mathcal{A}$, Max Turns $N$
\ENSURE Interaction History $\mathcal{H}$
\STATE \textbf{Initialise:} $\mathcal{H} \leftarrow \langle (P, \mathcal{M}(P)) \rangle$ \mycomment{Turn 0: Initial Code Generation}
\vspace{0.3em}
\FOR{$t = 1$ \TO $N-1$}
    \STATE $C_{prev} \leftarrow \proc{LastCode}(\mathcal{H})$; \quad $\tau \leftarrow \mathcal{A}[t]$
    \STATE $\Phi \leftarrow \proc{Shuffle}(\{i \in \mathcal{I} \mid \proc{Scope}(i) = \tau\})$
    \STATE $instr^* \leftarrow \text{NULL}$
    \FORALL{$i \in \Phi$} 
        \STATE \kw{if} $\texttt{Is\_Applicable}(C_{prev}, i)$ \kw{then} $instr^* \leftarrow i$; \kw{break}
    \ENDFOR

    \vspace{0.2em}
    \IF{$instr^* \neq \text{NULL}$}
        \STATE $C_t \leftarrow \mathcal{M}(\mathcal{H}, instr^*)$
        \STATE $\mathcal{I} \leftarrow \mathcal{I} \setminus \{instr^*\}$; \quad $\mathcal{H}.\proc{Append}((instr^*, C_t))$
    \ELSE
        \STATE $\mathcal{H}.\proc{Append}((\text{SKIP}, C_{prev}))$ \mycomment{Fallback: code maintained}
    \ENDIF
\ENDFOR
\RETURN $\mathcal{H}$
\end{algorithmic}
\end{algorithm}

\subsection{Evaluate Functional Correctness \& Instruction Adherence}
\label{sec:3.4}
In CodeChat-Eval, the quality of LLM-generated code is primarily evaluated by functional correctness. We also evaluate instruction adherence as a supporting diagnostic measure to distinguish whether failures arise because the model ignores a refinement instruction or because it applies the instruction while breaking functionality.

\subsubsection{Instruction Adherence}
\label{sec:3.4.2}

Following the practice of automated instruction adherence verification in related work\cite{mehralian2025codealignbench, zhou2023ifeval, yan2025codeif}, we implement a prompt-based approach to verify the instruction adherence of the LLM generated code using Chain-of-Thought\cite{wei2022cot} and few-shot prompting\cite{brown2020fewshot} to increase the accuracy of judgement. The prompt template is provided in supplementary material. According to the prompt, the underlying LLM is required to think step-by-step to 1) verify if the change between code before and code after adheres to the request in the provided instruction and 2) ignore the functional correctness when judging, then output the final verdict. If the code after is indeed revised from the code before as per the refinement instruction, it will be regarded as adhering to the instruction (\texttt{Adhere}), otherwise violating (\texttt{Violate}). We further verify the correctness of instruction adherence judgement in Section~\ref{sec:4.5}.


\subsubsection{Functional Correctness}
In CodeChat-Eval, we adopt the single-turn programming benchmarks as seed datasets. In each problem statement, a problem description, some examples of input-output (IO) pairs, and the function signature are provided. The LLM is expected to generate a code snippet that implements the function and follow the function name and IO format defined in signature. As the refinement instructions in follow-up turn are all functionality-preserving, the revised code is expected to follow the original function signature and realise the original functionality. Consequently, it remains feasible to use the original test suite from the programming benchmarks to verify the functional correctness of LLM-generated code in CodeChat-Eval.

To verify the functional correctness, the LLM output will first be sanitised to extract the function body from the plain text or markdown-formatted content. In this step, the predefined function signature will be used in pattern recognition and code extraction. Only if the LLM output follows the function signature and is successfully sanitised, and appears to be syntactically correct code that passes the whole test suite, can it be regarded as functionally correct (\texttt{Pass}). If the implementation is defective, or the LLM fails to follow the required function signature which interrupts the sanitation, the code will be regarded as functionally incorrect (\texttt{Fail}). 

To ensure accurate evaluation in CodeChat-Eval, it is important to make sure the follow-up instructions do not request to change 1) the function name, 2) the functionality description, and 3) the IO format. In CodeChat-Eval, as depicted in Section~\ref{sec:3.2}, we manually filtered out the instructions that are ambiguous, contradictory, the effect of which is incomparable, or conflicts with the aforementioned rules. This ensures the functional correctness judgement in CodeChat-Eval is reliable.

\subsection{Evaluation Metrics}
\label{sec:3.5}
CodeChat-Eval evaluates functional correctness using metrics tailored for multi-turn scenarios, and reports instruction adherence as a supporting diagnostic measure.

\subsubsection{Turn-Level Metrics}
In each evaluation session, code generated across all turns is verified against the corresponding test suite. To quantify functional correctness, we calculate the pass rate at each specific turn $T$. While the traditional \textbf{Pass@k} metric estimates the probability that at least one of $k$ generated solutions passes unit tests (typically $k \in \{1, 10, 100\}$) \cite{humaneval}, applying $k \ge 2$ in CodeChat-Eval is computationally prohibitive. A multi-turn evaluation session creates a branching factor that leads to an exponential explosion of the divergence space (e.g., a 10-turn evaluation session with $k=5$ requires executing $5^{10}$ branches). Consequently, we restrict our evaluation to the \textbf{Pass@1} setting. To ensure replicability, we employ greedy decoding for all evaluated LLMs. Similarly for instruction adherence, we define the \textbf{Instruction Adherence Rate (IAR)} as the proportion of generated solutions that strictly comply with the prompt's constraints at a given turn.

\subsubsection{Trend Detection via Mann-Kendall Test}
LLM should maintain consistent performance as the context grows. A downward trend indicates a degradation in capability during the multi-turn code refinement. To rigorously detect the trend over the turns, we employ the Mann-Kendall (MK) Test \cite{mann1945mk1,kendall1948mk2,kudrjavets2023mk3usage}. Unlike parametric approaches, the MK test does not assume a specific data distribution and is robust to noise. The MK Test statistic $S$ is calculated as $S = \sum_{k=1}^{n-1} \sum_{j=k+1}^{n} \text{sgn}(x_j - x_k)$, where $n$ is the sample size and $\text{sgn}(\cdot)$ is the sign function. To determine statistical significance, the standardised $Z_{MK}$ statistic is calculated as:
\begin{equation}
    Z_{MK} = \begin{cases} 
    \frac{S-1}{\sqrt{Var(S)}} & \text{if } S > 0 \\ 
    0 & \text{if } S = 0 \\ 
    \frac{S+1}{\sqrt{Var(S)}} & \text{if } S < 0 
    \end{cases}
\end{equation}
where $Var(S)$ is variance of $S$. We adopt a significance level of 0.05\cite{wohlin2012experimentation}. Such that if the computed $p$-value is less than $0.05$ ($|Z_{MK}| \geq 1.96$), we reject the null hypothesis ($H_0$) of no trend. This threshold implies a 95\% confidence level that the observed trend is deterministic and not a product of random chance. The direction of trend is given by the sign of $Z_{MK}$, a positive value ($Z_{MK} > 0$) indicates an upward trend, while a negative value ($Z_{MK} < 0$) indicates a downward trend. Both Pass@1 and IAR across turns are examined to rigorously detect the statistically significant trend.

\subsubsection{Mean Sustainable Turns (MST)}
Inspired by biostatistics\cite{royston2013restricted} and reliability engineering\cite{musa1975theorySRE,o2025PRE}, CodeChat-Eval uses the Mean Sustainable Turns (MST) metric to quantify the ability of LLMs to maintain functional correctness across evaluation sessions. For an evaluation session $i$, we define the Sustainable Turns ($ST_i$) as the count of consecutive turns where the LLM-generated code remains functionally correct starting from the initial turn (Turn 0). For example, if the code passes the test suite at Turn 0 and Turn 1 but becomes incorrect at Turn 2, then $ST_i = 2$. If the code at Turn 0 is already functionally incorrect, $ST_i = 0$. 
As the evaluation is limited to a maximum of $T_{max}$ turns, we record $ST_i = T_{max}$ if the LLM maintains functional correctness throughout the entire evaluation session. The MST is calculated as the arithmetic mean of $ST_i$ across all evaluation sessions:
\begin{equation}
    MST = \frac{1}{N} \sum_{i=1}^{N} ST_i
\end{equation}
where $N$ is the total number of evaluation sessions. Additionally, we use the notation MST@$T_{max}$ to denote the MST calculated with maximum turn number $T_{max}$. In CodeChat-Eval, all MST values are measured in a 10-turn window (one initial turn and nine follow-up turns) and are reported as MST@10.

\section{Experiments}
\label{sec:Exp}

\subsection{Seed Datasets and Evaluation Harness}
\label{sec:4.1}
In CodeChat-Eval, we adopt HumanEval\cite{humaneval} and MBPP\cite{MBPP} as the seed datasets. They provide problem statements, predefined function signatures, reference solutions, and default test suites. However, the default test suites suffer from limited test coverage and result in potential false positive results. To address this issue, we integrate EvalPlus\cite{liu2023evalplus} as our evaluation harness. As shown in Table~\ref{tab:sec4_1_seed}, EvalPlus augments the original test suites of HumanEval and MBPP with significantly extended test cases. Such that the functional correctness verification in CodeChat-Eval is strictly grounded.

\begin{table}[h]
    \centering
    \caption{Statistics of seed datasets. The Test Suites values represent the average number of test cases per task.}
    \label{tab:sec4_1_seed}
    \adjustbox{max width=\linewidth}{%
    \begin{tabular}{l|ccc}
    \toprule
         \textbf{Seed Dataset} & \textbf{Tasks} & \textbf{Original Tests} & \textbf{EvalPlus Tests} \\
         \midrule
         HumanEval & 164 & 10 & 784 \\
         MBPP & 378 & 3 & 105 \\
     \bottomrule
    \end{tabular}
    }
\end{table}

\subsection{Evaluated LLMs}
\label{sec:4.2}
CodeChat-Eval evaluates eight LLMs: Llama 3.1 8B \cite{llama3}, Llama 3.3 70B \cite{llama3}, Qwen 2.5 Coder (7B, 14B, 32B) \cite{qwen25coder}, DeepSeek-V3 (671B) \cite{deepseek}, GPT-5 Nano \cite{gpt5}, and GPT-5 \cite{gpt5}. The evaluated LLMs include six open-weight LLMs and two proprietary LLMs. The open-weight LLMs range from 7B to 671B parameters.

\subsection{Experiment Setup}
\label{sec:4.3}
All experiments are conducted on a computing cluster node with an Intel Xeon Platinum 8452Y CPU, 192GB RAM, and three NVIDIA H100 GPUs. The proprietary LLMs and DeepSeek-V3 LLM are accessed via their official APIs. The open-weight LLMs except DeepSeek-V3 are accessed via HuggingFace and use vLLM \cite{kwon2023vllm} as the inference engine. Greedy decoding (temperature = 0) is adopted for all evaluated LLMs to ensure reproducible evaluations.

The length of each evaluation session is fixed at 10 turns. The underlying LLMs for the instruction applicability check (Section~\ref{sec:3.3.1}) and the instruction adherence judgement (Section~\ref{sec:3.4.2}) are open-weight LLMs gpt-oss-20B and gpt-oss-120B\cite{agarwal2025gptoss} respectively. Instead of expensive proprietary LLMs\cite{mehralian2025codealignbench}, we adopt open-weight LLMs as underlying LLMs to ensure the long-term reproducibility of the evaluation framework and avoids the prohibitive API inference costs, ensuring the scalability of CodeChat-Eval. We manually verify the correctness of applicability check and instruction adherence judgement in Section \ref{sec:4.5}.

\subsection{Manual Verification}
\label{sec:4.5}
Before presenting the results, we manually verify the accuracy of automated judge of applicability (Section~\ref{sec:3.3.1}) and instruction adherence (Section~\ref{sec:3.4.2}). A total of 100 applicability check results and corresponding instruction adherence judgements are randomly extracted from the whole evaluation process on both datasets. Two authors independently labelled the binary correctness of each sample. For each applicability check result, if the two authors both regard the instruction is applicable to the code snippet, they label the result as \texttt{True}, otherwise \texttt{False}. For each instruction adherence result, if the two authors 1) consider the instruction indeed applicable to the code before; 2) agree that the revised code represents the intent of instruction, they label the instruction adherence judgement as \texttt{True}, otherwise \texttt{False}. Later, all disagreements are resolved through further in-depth discussion. According to the manual check results, the accuracy of applicability is 91\% with a substantial Cohen’s Kappa coefficient 0.709, the accuracy of instruction adherence judgement is 75\% with a good Cohen’s Kappa coefficient 0.639.

\section{Results}
\label{sec:Res}

\subsection*{\textbf{RQ1: How well do LLMs maintain functional correctness when applying the code refinement instructions?}}
\begin{figure}[!t]
    \centering
    \includesvg[width=1\linewidth]{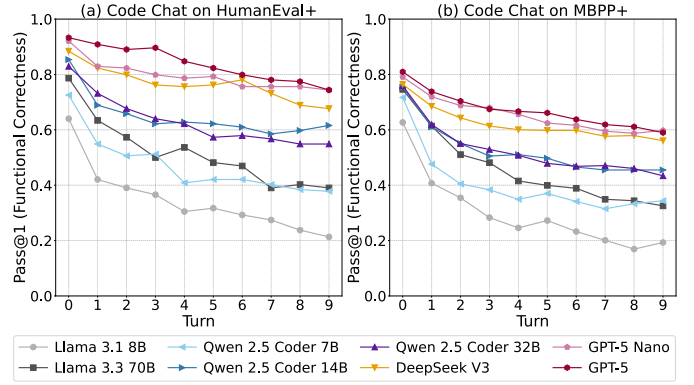}
    \caption{(RQ1) The decrease in functional correctness across multi-turn code refinement.}
    \label{fig:rq1_fc}
\end{figure}
\noindent\underline{\textbf{Results.}} 
\textbf{A decrease ranging from 19.2\% (GPT-5 Nano) to 69.2\% (Llama 3.1 8B) in functional correctness of the generated code is observed over the multi-turn refinement across all evaluated LLMs, the downward trend is verified as statistically significant on both MBPP and HumanEval.}

Fig~\ref{fig:rq1_fc} presents the functional correctness of LLM-generated code after the developer-preferred follow-up instructions are applied. A decrease ranging from 19.2\% (GPT-5 Nano) to 69.2\% (Llama 3.1 8B) in functional correctness of the generated code is observed over the multi-turn refinement across all evaluated LLMs. According to Table~\ref{tab:rq1_mk}, all LLMs exhibit statistically significant downward trend. Through the additional turns to the end, on average, the functional correctness of LLM-generated code on MBPP and HumanEval declines from 0.822 and 0.746, to 0.539 (-34.4\%) and 0.438 (-41.4\%) respectively, highlighting a significant quality deterioration of the generated code.

Among all evaluated LLMs, GPT-5 series models achieve the optimal performance to maintain functional correctness after the developer-preferred follow-up instructions are applied, while Llama models are among the worst. For GPT-5 Nano, we observe that the functional correctness decreases by 21.8\% when compared between the first and the last turns. For Llama 3.1 8B, the functional correctness decreases by 67.9\% on average across both datasets, indicating its weak capability to maintain the code functionally correct when revising the code. Although Llama 3.3 70B has a larger parameter size than Qwen 2.5 Coder 32B (+119\%), the correctness of Llama 3.3 70B generated code decreased by 53.4\% between the first and last turn across both datasets, which is 39.2\% higher than that of Qwen 2.5 Coder 32B (38.3\%). In longer contexts (turn 7-9), the performance Llama 3.3 70B and Qwen 2.5 Coder 7B become similar. The overall decrease of functional correctness of the code generated by the best-performing open-weight LLM, DeepSeek-V3, is 10.1\% higher than that of proprietary LLMs, suggesting a noticeable gap of coding capability between open-weight LLMs and proprietary LLMs. 

\begin{table}[!t]
    \centering
    \caption{(RQ1) Statistical analysis of functional correctness. $\Delta_{0\rightarrow9}$ denotes the percentage change in functional correctness from Turn 0 to Turn 9. $p_{value}$ indicates the significance of trend based on Mann-Kendall test. Bold values marked with * indicate a statistically significant trend ($p_{value} < 0.05$). $\downarrow$ indicates a downward trend.}
    \label{tab:rq1_mk}
    \adjustbox{max width=\linewidth}{%
    \begin{tabular}{l|ccc|ccc}
        \toprule
        \multirow{2}{*}{\textbf{Model}} & \multicolumn{3}{c|}{\textbf{HumanEval}} & \multicolumn{3}{c}{\textbf{MBPP}} \\
        \cmidrule(lr){2-4} \cmidrule(lr){5-7}
         & \textbf{$\Delta_{0\rightarrow9}$} & \textbf{$p_{value}$} & \textbf{Trend} & \textbf{$\Delta_{0\rightarrow9}$} & \textbf{$p_{value}$} & \textbf{Trend} \\
        \midrule
        Llama 3.1 8B & -66.67\% & \textbf{0.000}$^*$ & $\downarrow$ & -69.20\% & \textbf{0.000}$^*$ & $\downarrow$ \\
        Llama 3.3 70B & -50.39\% & \textbf{0.000}$^*$ & $\downarrow$ & -56.38\% & \textbf{0.000}$^*$ & $\downarrow$ \\
        Qwen 2.5 Coder 7B & -47.90\% & \textbf{0.001}$^*$ & $\downarrow$ & -52.03\% & \textbf{0.002}$^*$ & $\downarrow$ \\
        Qwen 2.5 Coder 14B & -27.86\% & \textbf{0.003}$^*$ & $\downarrow$ & -39.65\% & \textbf{0.000}$^*$ & $\downarrow$ \\
        Qwen 2.5 Coder 32B & -33.82\% & \textbf{0.000}$^*$ & $\downarrow$ & -42.86\% & \textbf{0.000}$^*$ & $\downarrow$ \\
        DeepSeek-V3 & -23.45\% & \textbf{0.002}$^*$ & $\downarrow$ & -26.64\% & \textbf{0.000}$^*$ & $\downarrow$ \\
        GPT-5 Nano & -19.21\% & \textbf{0.000}$^*$ & $\downarrow$ & -24.41\% & \textbf{0.000}$^*$ & $\downarrow$ \\
        GPT-5 & -20.26\% & \textbf{0.000}$^*$ & $\downarrow$ & -27.12\% & \textbf{0.000}$^*$ & $\downarrow$ \\
        \bottomrule
    \end{tabular}
    }
\end{table}

\begin{table}[!t]
    \centering
    \caption{(RQ1) The Mean Sustainable Turns (MST@10) of evaluated LLMs.}
    \label{tab:rq1_MST}
    \adjustbox{max width=\linewidth}{%
        \begin{tabular}{l|r|r|r}
        \toprule
           \textbf{LLM} & \textbf{HumanEval} & \textbf{MBPP} & \textbf{Mean} \\
        \midrule
            Llama 3.1 8B & 2.439 & 2.177 & 2.308 \\
            Llama 3.3 70B & 3.768 & 3.479 & 3.624 \\
            Qwen 2.5 Coder 7B & 2.994 & 2.931 & 2.963 \\
            Qwen 2.5 Coder 14B & 4.970 & 3.717 & 4.343 \\
            Qwen 2.5 Coder 32B & 5.091 & 3.767 & 4.429 \\
            Deepseek V3 & 6.762 & 5.310 & 6.036 \\
            GPT 5 nano & 7.037 & 5.810 & 6.423 \\
            GPT 5 & 7.762 & 6.026 & 6.894 \\
        \midrule
            Mean & 5.103 & 4.152 & 4.628 \\
        \bottomrule
        \end{tabular}}
\end{table}

MST serves as a unified metric that incorporates both the initial functional correctness and the performance decay over successive turns, representing the overall functional correctness of an LLM in multi-turn code refinement. According to the results in Table~\ref{tab:rq1_MST}, GPT-5 achieves the highest MST@10 of 6.894 across both datasets. While GPT-5 Nano exhibits a lower rate of degradation, its functional correctness is generally lower than that of GPT-5 in most turns. Consequently, GPT-5 exhibit higher overall functional correctness, as reflected by its higher MST@10 value compared to GPT-5 Nano (+7.3\%). Conversely, the average MST@10 of Llama 3.1 8B is 2.308, indicating that this model typically fails to maintain correctness after the second follow-up turn. This suggests that the coding capability of Llama 3.1 8B is limited in multi-turn code refinement. The MST@10 values of the evaluated LLMs correspond with their overall performance in CodeChat-Eval, where a higher Pass@1 in the initial turn and a lower degradation rate together result in a higher MST@10. Thus, MST is a representative metric for multi-turn code refinement evaluation. The sensitivity of MST to $T_{max}$ is further discussed in Section~\ref{sec:dis.sst}.

\begin{tcolorbox}[colback=gray!10, colframe=black, boxrule=0.5pt]
\textbf{Answer to RQ1:} A decrease ranging from 19.2\% (GPT-5 Nano) to 69.2\% (Llama 3.1 8B) in functional correctness of the generated code is observed over the multi-turn refinement across all evaluated LLMs, the downward trend is verified as statistically significant on both MBPP and HumanEval, indicating the evaluated LLMs struggle to maintain functional correctness in multi-turn code refinement.
\end{tcolorbox}

\subsection*{\textbf{RQ2: Which types of code refinement instructions have the largest impact on the functional correctness of the LLM-generated code?}}

\noindent\underline{\textbf{Results.}} 
\textbf{Among the three instruction scopes, \scopeSemantic{Semantic} instructions, which require changes to the underlying logic or execution profile, cause the largest decrease in functional correctness, resulting in a 7.3\% to 34.0\% regression rate. Among the three change operations, \changeAdd{Add} instructions, which request additional content, features, or code patterns, cause the largest decrease in functional correctness, resulting in a 6.3\% to 27.9\% regression rate.}

\subsubsection*{RQ2.1 \#Instruction Scope}
\begin{figure}[!t]
    \centering
    \includesvg[width=1\linewidth]{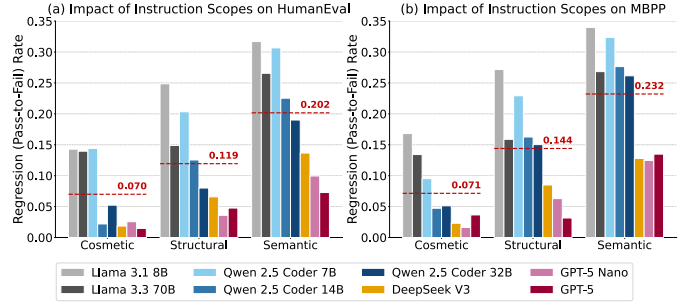}
    \caption{(RQ2.1) The regression rate of different scopes.}
    \label{fig:rq2_scope}
\end{figure}

To evaluate the influence of different instruction scopes, we calculate the regression (Pass-to-Fail) rate, which represents the proportion of previously correct code that fails after refinement. As shown in Fig.~\ref{fig:rq2_scope}, \scopeSemantic{Semantic} instructions, which involve changes to the underlying logic or execution profile while preserving the original interface and intended behaviour, correspond to the highest regression rates on HumanEval (0.202) and MBPP (0.232). Among the evaluated LLMs, Llama 3.1 8B exhibits the highest regression rates. Overall, \scopeSemantic{Semantic} instructions are consistently associated with the most substantial decreases in functional correctness, recording regression rates between 7.3\% and 34.0\%.
\begin{figure}[!t]
    \centering
    \includesvg[width=1\linewidth]{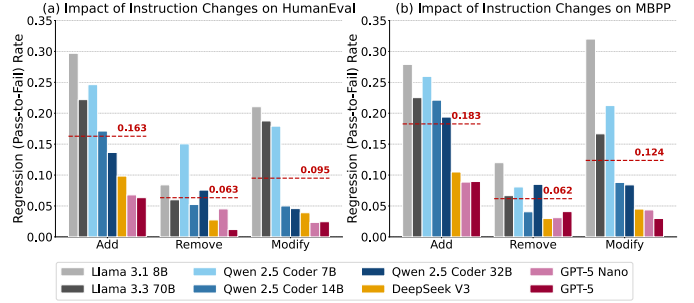}
    \caption{(RQ2.2) The regression rate of different changes.}
    \label{fig:rq2_change}
\end{figure}

\subsubsection*{RQ2.2 \#Instruction Change}
Similarly, we calculate regression rates to assess the influence of different change operations. As shown in Fig.~\ref{fig:rq2_change}, \changeAdd{Add} instructions, which ask the model to introduce additional content, features, or code patterns, are associated with the highest regression rates, averaging 0.163 on HumanEval and 0.183 on MBPP. Overall, applying \changeAdd{Add} instructions consistently corresponds to a degradation in functional correctness, with regression rates ranging from 6.3\% to 27.9\%.

\begin{tcolorbox}[colback=gray!10, colframe=black, boxrule=0.5pt]
\textbf{Answer to RQ2:} Instructions that require changes to the underlying logic or execution profile (\scopeSemantic{Semantic}, decreased functional correctness by 21\%) and instructions that request additional content, features, or code patterns (\changeAdd{Add}, decreased functional correctness by 17\%) have the largest negative impact on the functional correctness of the LLM-generated code.
\end{tcolorbox}

\section{Discussion}

\subsection{Sensitivity of MST to \texorpdfstring{$T_{max}$}{Tmax}}
\label{sec:dis.sst}
We conducted a sensitivity analysis to evaluate how the maximum turn count ($T_{max}$) influences the MST metric. Across all evaluation sessions for all LLMs, only 6 out of 39,024 ($542\times8\times9$) turns (0.015\%) contained \textit{SKIP} flags. This indicates a negligible impact on the MST calculation, which treats a skipped turn as maintaining the status of the previous turn. 

To assess sensitivity, we compared $T_{max}$ values ranging from 2 to 10 with a step of 2. By virtually truncating the evaluation sessions at different $T_{max}$ points, we recalculated the MST as shown in Table~\ref{tab:MST_sens}. Although the absolute MST value naturally increases as the upper bound $T_{max}$ rises, the relative rate of change ($\Delta$) diminishes significantly for most models. This suggests a trend towards stability in capturing the model's failure characteristics. While a larger $T_{max}$ could further refine the metric, the benefit decreases as $T_{max}$ grows. Since $T_{max}=10$ covers approximately 95\% of real-world developer-LLM code refinement dialogues, the setting in CodeChat-Eval is considered both appropriate and sufficient.

\begin{table}[h!]
    \centering
    \caption{(Discussion A) Sensitivity Analysis of MST across Different Interaction Turns ($T_{max}$). The percentage in parentheses denotes the relative change compared to the previous column ($\Delta = \frac{\text{Current} - \text{Previous}}{\text{Previous}}$). Bold values indicate the lowest relative change across the row.}
    \label{tab:MST_sens}
    \adjustbox{max width=\linewidth}{%
    \begin{tabular}{ll|rrrrr}
    \toprule
    \multirow{2}{*}{\textbf{Dataset}} & \multirow{2}{*}{\textbf{Model}} & \multicolumn{5}{c}{\textbf{MST@$T_{max}$}} \\
    \cmidrule(lr){3-7}
     & & \textbf{2} & \textbf{4} & \textbf{6} & \textbf{8} & \textbf{10} \\
    \midrule
    \multirow{8}{*}{\rotatebox[origin=c]{90}{HumanEval}}
    & Llama 3.1 8B & 1.055 & 1.665 (+57.8\%) & 2.043 (+22.7\%) & 2.293 (+12.2\%) & \textbf{2.439 (+6.4\%)} \\
    & Llama 3.3 70B & 1.409 & 2.335 (+65.8\%) & 2.988 (+27.9\%) & 3.457 (+15.7\%) & \textbf{3.768 (+9.0\%)} \\
    & Qwen 2.5 Coder 7B & 1.244 & 1.988 (+59.8\%) & 2.433 (+22.4\%) & 2.756 (+13.3\%) & \textbf{2.994 (+8.6\%)} \\
    & Qwen 2.5 Coder 14B & 1.543 & 2.701 (+75.1\%) & 3.628 (+34.3\%) & 4.372 (+20.5\%) & \textbf{4.970 (+13.7\%)} \\
    & Qwen 2.5 Coder 32B & 1.549 & 2.732 (+76.4\%) & 3.689 (+35.0\%) & 4.470 (+21.2\%) & \textbf{5.091 (+13.9\%)} \\
    & DeepSeek V3 & 1.695 & 3.201 (+88.8\%) & 4.543 (+41.9\%) & 5.756 (+26.7\%) & \textbf{6.762 (+17.5\%)} \\
    & GPT-5 Nano & 1.744 & 3.256 (+86.7\%) & 4.634 (+42.3\%) & 5.872 (+26.7\%) & \textbf{7.037 (+19.8\%)} \\
    & GPT-5 & 1.835 & 3.537 (+92.7\%) & 5.067 (+43.3\%) & 6.470 (+27.7\%) & \textbf{7.762 (+20.0\%)} \\
    \midrule
    \multirow{8}{*}{\rotatebox[origin=c]{90}{MBPP}}
    & Llama 3.1 8B & 1.029 & 1.566 (+52.2\%) & 1.868 (+19.3\%) & 2.058 (+10.2\%) & \textbf{2.177 (+5.8\%)} \\
    & Llama 3.3 70B & 1.339 & 2.188 (+63.4\%) & 2.759 (+26.1\%) & 3.172 (+15.0\%) & \textbf{3.479 (+9.7\%)} \\
    & Qwen 2.5 Coder 7B & 1.180 & 1.823 (+54.5\%) & 2.283 (+25.3\%) & 2.646 (+15.9\%) & \textbf{2.931 (+10.8\%)} \\
    & Qwen 2.5 Coder 14B & 1.352 & 2.238 (+65.6\%) & 2.868 (+28.1\%) & 3.341 (+16.5\%) & \textbf{3.717 (+11.2\%)} \\
    & Qwen 2.5 Coder 32B & 1.357 & 2.238 (+64.9\%) & 2.873 (+28.4\%) & 3.376 (+17.5\%) & \textbf{3.767 (+11.6\%)} \\
    & DeepSeek V3 & 1.434 & 2.585 (+80.3\%) & 3.571 (+38.2\%) & 4.476 (+25.3\%) & \textbf{5.310 (+18.6\%)} \\
    & GPT-5 Nano & 1.505 & 2.788 (+85.2\%) & 3.905 (+40.0\%) & 4.897 (+25.4\%) & \textbf{5.810 (+18.6\%)} \\
    & GPT-5 & 1.534 & 2.833 (+84.7\%) & 3.997 (+41.1\%) & 5.058 (+26.5\%) & \textbf{6.026 (+19.1\%)} \\
    \bottomrule
    \end{tabular}
    }
\end{table}

\subsection{Instruction Adherence and Functional Correctness}
\begin{figure}[h!]
    \centering
    \includesvg[width=1\linewidth]{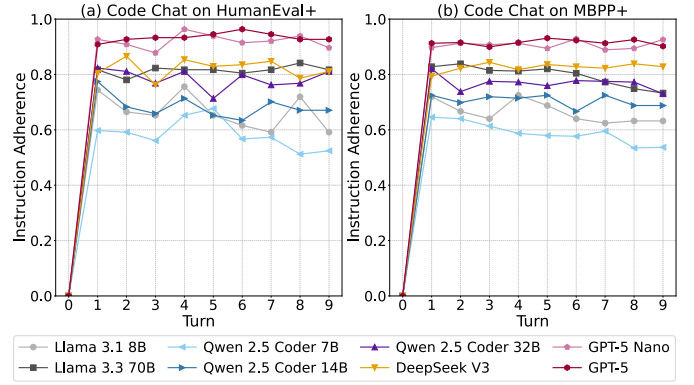}
    \caption{Instruction adherence rate across multi-turn code refinement.}
    \label{fig:disc_ia}
\end{figure}

\begin{table}[h!]
    \centering
    \caption{Statistical analysis of instruction adherence across multi-turn code refinement (Turns 1--9). $\Delta_{1\rightarrow9}$ denotes the percentage change in instruction adherence from Turn 1 to Turn 9. $p_{value}$ indicates the significance of the trend based on the Mann-Kendall test. Bold values marked with $^*$ indicate a statistically significant trend ($p_{value} < 0.05$), and $\downarrow$ indicates a downward trend.}
    \label{tab:disc_ia}
    \adjustbox{max width=\linewidth}{%
    \begin{tabular}{l|ccc|ccc}
        \toprule
        \multirow{2}{*}{\textbf{Model}} & \multicolumn{3}{c|}{\textbf{HumanEval}} & \multicolumn{3}{c}{\textbf{MBPP}} \\
        \cmidrule(lr){2-4} \cmidrule(lr){5-7}
         & \textbf{$\Delta_{1\rightarrow9}$} & \textbf{$p_{value}$} & \textbf{Trend} & \textbf{$\Delta_{1\rightarrow9}$} & \textbf{$p_{value}$} & \textbf{Trend} \\
        \midrule
        Llama 3.1 8B & -20.49\% & 0.073 & - & -12.13\% & \textbf{0.045}$^*$ & $\downarrow$ \\
        Llama 3.3 70B & 0.00\% & 0.565 & - & -11.50\% & \textbf{0.002}$^*$ & $\downarrow$ \\
        Qwen 2.5 Coder 7B & -12.24\% & 0.175 & - & -16.80\% & \textbf{0.005}$^*$ & $\downarrow$ \\
        Qwen 2.5 Coder 14B & -13.39\% & 0.295 & - & -5.11\% & 0.336 & - \\
        Qwen 2.5 Coder 32B & -1.48\% & 0.336 & - & -10.97\% & 0.343 & - \\
        DeepSeek-V3 & 0.76\% & 0.754 & - & 4.33\% & 0.343 & - \\
        GPT-5 Nano & -3.29\% & 1.000 & - & 3.24\% & 1.000 & - \\
        GPT-5 & 2.01\% & 0.389 & - & -1.16\% & 0.752 & - \\
        \bottomrule
    \end{tabular}
    }
\end{table}

We use instruction adherence as a supporting diagnostic analysis to examine whether the functional correctness degradation observed in RQ1 is simply caused by models ignoring refinement instructions. Since Turn 0 involves code generation from an initial requirement without specific refinement constraints, instruction adherence is evaluated from Turn 1 to Turn 9. As shown in Fig.~\ref{fig:disc_ia} and Table~\ref{tab:disc_ia}, the change in instruction adherence rate varies from -20.5\% (Llama 3.1 8B) to +4.3\% (DeepSeek-V3). On HumanEval, none of the evaluated LLMs exhibit a statistically significant degradation in instruction adherence. On MBPP, five out of eight evaluated LLMs do not exhibit a statistically significant downward trend.

To further understand the relationship between functional correctness and instruction adherence, we aggregate the number of turns that the LLM-generated code is functionally correct and instruction adhered (or not). Among the total 20623 functionally correct (\texttt{Pass}) turns, the numbers of instruction adhered turns and instruction violated turns are 16668 and 3955 respectively. Among a total of 18395 functionally incorrect (\texttt{Fail}) turns, the numbers of instruction adhered turns and instruction violated turns are 13497 and 4898 respectively. The Phi coefficient\cite{fleiss2013statistical} between functional correctness and instruction adherence is 0.089, which is below 0.1\cite{cohen2013statistical}, indicating a negligible correlation. Consequently, successfully adhering to a refinement instruction does not guarantee that the revised code will maintain functional correctness. This supports our main finding that functional correctness degradation is not merely an instruction-adherence failure; it reflects a distinct weakness of LLMs in preserving behaviour during multi-turn code refinement.

\subsection{Self-Correction During Multi-Turn Code Refinement}

\begin{figure}[!t]
    \centering
    \includesvg[width=1\linewidth]{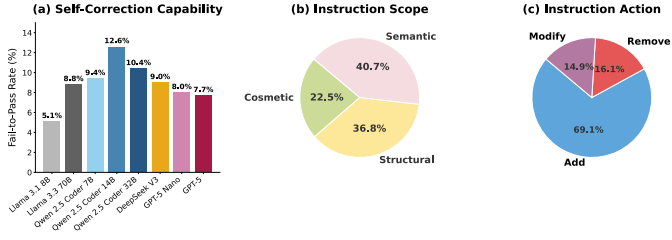}
    \caption{(Discussion C) The self-correction in multi-turn code refinement. (a) The self-correction rate across different LLMs, and self-correction occurs with instructions of different (b) scopes, and (c) changes.}
    \label{fig:sec64}
\end{figure}

We evaluate the self-correction capability of LLMs using the Fail-to-Pass rate, which measures the ratio of revised functionally incorrect code in previous turns to be functionally correct and pass the test suite. The results are shown in Fig.~\ref{fig:sec64}(a). Among all evaluated LLMs, Qwen 2.5 Coder 14B achieves the highest self-correction rate (12.6\%), whereas both the largest and the smallest evaluated models exhibit lower rates. This observation indicates a potential non-linear relationship between model scale and self-correction capability, which warrants further investigation in future work.

The \scopeSemantic{Semantic} typed and \changeAdd{Add} typed instructions are more likely to trigger the self-correction. As the refinement instructions are all functionality-preserving, it is anticipated that the self-correction is a potential side-effect of the code revision process. The \scopeSemantic{Semantic} scope and \changeAdd{Add} change typically require modifications to underlying logic, execution profile, or additional code patterns, which might patch the originally ignored edge cases or defective implementation.

\section{Threats to Validity}
\label{sec:TtV}

\subsection{Internal Validity}
\textbf{Accuracy of prompt-based evaluation:} CodeChat-Eval uses a prompt-based approach based on open-weight LLMs to evaluate instruction applicability and instruction adherence. This may lead to inaccuracy in judgement. To mitigate inaccuracies, CodeChat-Eval uses CoT and few-shot prompting (Section~\ref{sec:3.4}). As detailed in Section~\ref{sec:4.5}, manual verification confirms the accuracy of the adopted approach.

\textbf{Instruction filtering and test suite execution:} A small proportion (6.51\%) of code refinement instructions conflict with the fixed function signatures of the test suites. To ensure the test suites accurately evaluate functional correctness, we filter out these conflicting instructions. Although this exclusion slightly reduces the breadth of the instruction set, it ensures the failed test suites are caused by functionally incorrect code rather than disrupted testing harness.

\textbf{Data contamination:} CodeChat-Eval uses HumanEval and MBPP as seed datasets. Data contamination is the phenomenon that the evaluated LLMs may have encountered these evaluation datasets during training\cite{guo2025mortar}. However, CodeChat-Eval uses the dynamic generation algorithm (AGDIS) to produce evaluation sessions. These dynamically generated evaluation sessions are unlikely to exist in the training corpora, thereby mitigating data contamination in multi-turn code refinement evaluation.

\subsection{External Validity}
\textbf{Complexity of multi-turn dialogues:} In real-world scenarios, developers may use composite instructions containing multiple code refinement requests. Additionally, real-world multi-turn dialogues may exceed the 10-turn limit ($T_{max}=10$) used in CodeChat-Eval evaluation sessions. CodeChat-Eval evaluates single code refinement instructions to isolate the impact of different instruction scopes and change operations. The setting of 10-turn ($T_{max}=10$) covers 95\% of real-world multi-turn dialogues. Furthermore, CodeChat-Eval currently only evaluates Python code. However, the code refinement instructions and the evaluation framework are potentially applicable to other programming languages. Future work will extend CodeChat-Eval to composite instructions and additional programming languages.

\textbf{Generalisation:} CodeChat-Eval curates code refinement instructions from a specific dataset \cite{mehralian2025codealignbench}. These instructions may not represent all possible code refinement instructions used in real-world software development. However, CodeChat-Eval classifies these instructions using a general instruction taxonomy (Section~\ref{sec:3.1}). This taxonomy ensures that the evaluation of instruction scopes and change operations generalises to broader multi-turn code refinement tasks.

\section{Conclusion}
\label{sec:Conclu}
This paper introduces CodeChat-Eval, an evaluation framework designed to evaluate LLMs in multi-turn code refinement. CodeChat-Eval uses the Agenda-Guided Dynamic Instruction Selection (AGDIS) algorithm to generate dynamic evaluation sessions. By dynamically applying functionality-preserving code refinement instructions, CodeChat-Eval evaluates how LLMs maintain functional correctness across multi-turn code refinement.

The empirical study on eight LLMs indicates that the functional correctness of the generated code exhibits a statistically significant decrease across a 10-turn evaluation session. Specifically, instructions that require changes to the underlying logic or execution profile (\scopeSemantic{Semantic}, decreased functional correctness by 21\%) and instructions that request additional content, features, or code patterns (\changeAdd{Add}, decreased functional correctness by 17\%) have the largest negative impact on the functional correctness of the LLM-generated code. The proposed Mean Sustainable Turns (MST) metric corresponds with the overall functional correctness of LLM-generated code in multi-turn refinement. Since existing single-turn benchmarks cannot expose this functional correctness degradation, our findings motivate multi-turn refinement benchmarks beyond single-turn code generation.

Future work will analyse real-world multi-turn dialogues to refine the evaluation agenda. Additionally, we will evaluate composite code refinement instructions and extend CodeChat-Eval to support additional programming languages.

\section{Data Availability}
The replication package, prompts and results are publicly accessible at https://zenodo.org/records/18893780.

\bibliographystyle{IEEEtran}
\bibliography{ref}

\end{document}